\documentclass[nopacs,epj]{svjour}
\usepackage{graphicx}

\usepackage{amssymb}
\usepackage{amsmath}

\usepackage{color}

\usepackage[colorlinks]{hyperref}

\AtBeginDocument{%
    \hypersetup{
      urlcolor     = blue, 
      linkcolor    = black, 
      citecolor   = blue, 
      pdfborder={0 0 0},
      urlbordercolor={1 1 1},
      citebordercolor={1 1 1}
    }
} 

\usepackage{multirow}

\begin{document}

\hyphenation{coun-ter cor-res-pon-din-gly e-xam-ple
co-o-pe-ra-tion ex-pe-ri-men-tal pum-ping ve-ri-fi-ca-tion ve-ri-fi-ca-tions ha-ving pro-ba-bi-li-ty}

\title{Spying on photons with photons: quantum interference and information}


\author{Stefan~Ataman\inst{1}
}                     

\institute{Extreme Light Infrastructure - Nuclear Physics
(ELI-NP), 30 Reactorului Street, 077125 Magurele, 
Romania \email{stefan.ataman@eli-np.ro} }

\date{Received: date / Revised version: date}
%
\abstract{The quest to have both which-path knowledge and
interference fringes in a double-slit experiment dates back to the
inception of quantum mechanics (QM) and to the famous
Einstein-Bohr debates. In this paper we propose and discuss an
experiment able to spy on one photon's path with another photon. We modify the quantum state inside the interferometer as opposed to the traditional physical modification of the ``wave-like'' or ``particle-like'' experimental setup.
We are able to show that it is the ability to harvest or not which-path information that finally
limits the visibility of the interference pattern and not the ``wave-like'' or ``particle-like'' experimental setups. Remarkably, a full ``particle-like'' experimental setup is able to show interference fringes with $100\:\%$ visibility if the quantum state is carefully engineered.
%
\PACS{
      {PACS-key}{describing text of that key}   \and
      {PACS-key}{describing text of that key}
     } 
} 
%


\maketitle
%

\section{Introduction}

Discussions about the wave-particle duality and the
counter-intuitive features of quantum mechanics started almost a
century ago with the famous Einstein-Bohr debates \cite{Boh49}.

It was generally accepted that the loss of interference in a
two-slit experiment is a consequence of the perturbation induced
by the measuring device \cite{Fey65}, perturbation lower bounded
by the Heisenberg uncertainty principle. Bohr \cite{Boh28} was the
first to realize that Quantum Mechanics is more subtle than this
``observation caused disturbance induces the uncertainty'' dictum
and his Complementary Principle refined the so-called
wave-particle duality to a new level. The detailed quantum
mechanical treatment of Einstein's recoiling slit experiment was
done by Wootters and Zurek \cite{Woo79}. The authors also discuss
the partial which-path knowledge and the reduced contrast of the
interference fringes, thus bringing a quantitative discussion to
Bohr's complementarity principle. The inequality $K^2+V^2\leq1$
(where $K$ quantifies the which-path
information and $V$ the interference fringe visibility) was
successively refined and discussed by many authors
\cite{Woo79,Dur00,Gre88,Eng95,Eng96}.

A couple of decades after Bohr's proposal, Wheeler
\cite{Whe78,Whe83} introduced the idea of delayed-choice
experiment. In his original \emph{Gedankenexperiment} one could
decide to remove or not the second beam splitter in a Mach-Zehnder
interferometer (MZI) after the light quantum left the first beam
splitter. This delayed choice decides on the ``wave-like'' or
``particle-like'' phenomenon that is measured. And if this
decision is space-like separated with the passage of the light
quantum in the first beam splitter, no causal link could exist
between these two events. It took several decades until Jacques
\emph{et al.} \cite{Jac07,Jac08} fully tested Wheeler's original
delayed choice proposal. However, physicists had little doubts of
its outcome and previous experiments tested equivalent schemes.
Using extremely attenuated laser light, Hellmuth, Walther, Zajonc
and W. Schleich \cite{Hel87} performed such an experiment. Their
results show ``no observable difference between normal and
delayed-choice modes of operation'', as predicted by quantum
mechanics. Two years later, using single-photon states Baldzuhn,
Mohler and Martienssen \cite{Bal89} performed the same experiment
and arrived at a similar conclusion. Complementarity was also
considered with atoms scattering light \cite{Cha95,Eic93,Dur98} up
to the quantum classical boundary \cite{Ber01}.

A new twist in this experiment arrived with the proposal of
Ionicioiu and Terno \cite{Ion11}, where -- in a truly quantum
mechanical style -- the second beam splitter is in a superposition
of being and not being inserted. One can therefore morph between ``wave'' and ``particle'' behavior in a continuous
manner. Experimental verifications followed by Kaiser et al.
\cite{Kai12} and Peruzzo et al. \cite{Per12}.

Meanwhile the new idea of \emph{Quantum Eraser} was proposed by
Scully and Dr\"uhl \cite{Scu82}. In fact, it was possible to
``erase'' a which path information and -- surprisingly -- revive
the interference fringes previously washed out by the which-path
markers \cite{Scu91,Eng99,Scu98}. The original experiment with emitting
atoms and micromaser cavities was deemed too difficult to
implement, therefore focus was set on equivalent optical
implementations \cite{Wal02,Hil98}. Even a do-it-yourself quantum
eraser has been suggested \cite{Hil98}. For a review of delayed-choice and quantum eraser proposals and experiments see reference \cite{Ma16}.

Using a bi-photon state to realize a quantum eraser seems to be
first proposed by Ou \cite{Ou97} by using two parametric
down-converters and the ``phase memory'' of the pumping beam
\cite{Ou90}. Recently, in \cite{Heu15} three non-linear crystals
where employed to show the complementarity between the visibility
of interference fringes and the which-path knowledge, proving that
this subject is still an active research topic.

In this paper we revisit and modify the experiment proposed in reference
\cite{Ata14b}. The main idea -- spying on photons inside a MZI
with photons leaking out of it -- is thoroughly discussed and taken one step further. We are able to show
that one does not need a quantum eraser to recover interference:
if the quantum state inside the MZI is carefully engineered so
that no which-path information can be inferred from the ``leaked''
photon, interference fringes are always recovered, even in a full ``particle-like'' experimental setting. Similar to
previous proposals \cite{Ion11,Kai12}, a continuous morphing
between wave-like and particle-like behavior is also possible. This time, however, we have a \emph{fixed} experimental setup and we morph between particle-like and wave-like behavior by modifying the input state vector.

The paper is organized as follows. The classic subject of quantum interference and path information with or without
delayed choice in a MZI is given in Section
\ref{sec:interference_MZI}. The idea of spying on photons with
photons is introduced and discussed in Section
\ref{sec:spying_on_photons}. The input state is modified in Section
\ref{sec:spying_modif_input_state} so that
the which-path information obtainable from the inner MZI can be
varied from zero to maximal. Finally conclusions are
drawn in Section \ref{conclusions}.

\section{Interference and which-path information with a Mach-Zehnder interferometer}
\label{sec:interference_MZI} 
\begin{figure}
\centering
\includegraphics[width=2in]{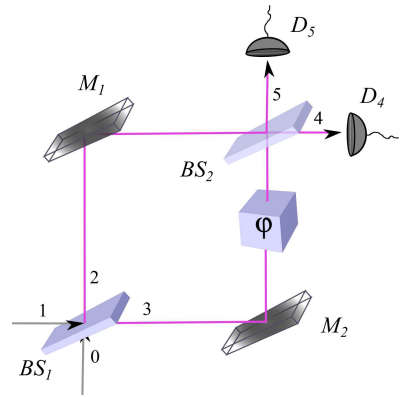}
\caption{The Mach-Zehnder interferometer. The phase shift $\varphi$ models a
voluntarily introduced path length difference.
Wheeler's delayed choice can be added as suggested by the semi-transparent
beam splitter $\text{BS}_2$.} \label{fig:Simple_MZIs_delayed_choice}
\end{figure}

A Mach-Zehnder interferometer is
depicted in Fig.~\ref{fig:Simple_MZIs_delayed_choice}. It is
composed of the beam splitters $\text{BS}_1$ and $\text{BS}_2$,
together with the mirrors $\text{M}_1$ and $\text{M}_2$. The beam
splitters are assumed identical and are characterized by the
transmission (reflection), coefficients $T$ ($R$). Unitarity
constrains imply $\vert{T}\vert^2+\vert{R}\vert^2=1$ and
$RT^*+TR^*=0$ \cite{Lou03}.

The delay $\varphi$ models the path length difference of the
interferometer.
Detectors $D_{4}$ and $D_{5}$ are placed at the two outputs of
beam splitter $\text{BS}_2$. Throughout this paper, we shall
assume monochromatic light and ideal photo-detectors.

When dealing with a balanced (50/50) beam splitter, we shall use
$T=1/\sqrt{2}$ and $R=i/\sqrt{2}$.

If we apply a single photon Fock state at the input $1$ of our
interferometer, we can write the input state as
\begin{equation}
\label{eq:MZI_single_photon_input}
\vert\psi_{in}\rangle=\vert1_10_0\rangle=\hat{a}_1^\dagger\vert0\rangle
\end{equation}
where $\hat{a}_k^\dagger$ denotes the
creation operator at the port $k$. After the first beam splitter,
the state vector can be written as \cite{Lou03}
\begin{equation}
\label{eq:MZI_after_first_BS}
\vert\psi_{23}\rangle=R\vert1_20_3\rangle+T\vert0_21_3\rangle
\end{equation}
This is an entangled state \cite{Enk05}. Our photon is no more in
a definitive path, it is in a coherent superposition of being in
both the upper and lower paths (with probabilities
$\vert{R}\vert^2$ and, respectively, $\vert{T}\vert^2$). Using the
field operator transformation (see e. g. \cite{Lou03},
\cite{Ata14a})
\begin{equation}
\hat{a}_1^\dagger={\color{black}TR\left(1+e^{i\varphi}\right)}\hat{a}_4^\dagger
+{\color{black}\left(T^2e^{i\varphi}+R^2\right)}\hat{a}_5^\dagger
\end{equation}
one can find right away the output state vector
\begin{equation}
\label{eq:MZI_final_gen_state} \vert\psi_{out}\rangle
=TR\left(1+e^{i\varphi}\right)\vert1_40_5\rangle
+\left(T^2e^{i\varphi}+R^2\right)\vert0_41_5\rangle
\end{equation}
and if both beam splitters are balanced (50/50), we get
\begin{equation}
\label{eq:psi_out_MZI_balanced} \vert\psi_{out}\rangle
=\cos\left(\varphi/2\right)\vert1_40_5\rangle
+\sin\left(\varphi/2\right)\vert0_41_5\rangle
\end{equation}
where we neglected a common phase factor. The probability of
photo-detection at, say, detector $D_4$ is simply
$P_4=\vert\langle1_40_5\vert\psi_{out}\rangle\vert^2$ yielding
\begin{equation}
P_4=\cos^2\left(\varphi/2\right)=\frac{1+\cos\left(\varphi\right)}{2}
\end{equation}
We recognize here the well-known interference fringes, having
$100$ \% visibility. We had no which-path knowledge whatsoever,
therefore following Feynman's rules the (complex) amplitudes
corresponding to the two paths are added up, yielding this
interference phenomenon.

Starting from equations
\eqref{eq:MZI_single_photon_input}--\eqref{eq:MZI_final_gen_state}
we can define the which-path information as
\begin{equation}
K=\big\vert\vert{T}\vert^2-\vert{R}\vert^2\big\vert
\end{equation}
and the
interference fringe visibility 
\begin{equation}V=2\vert{T}\vert\vert{R}\vert.
\end{equation}
and a short computation yields immediately $K^2+V^2=1$.

As a final remark, equation \eqref{eq:MZI_after_first_BS} states
that our single photon is in a coherent superposition of being in
both arms at the same time. This is not just a metaphor and one
could easily convince oneself by changing the input state so that
the state vector after the beam splitter $\text{BS}_1$ becomes e.
g.
\begin{equation}
\label{eq:MZI_after_first_BS_onepath}
\vert\psi_{23}\rangle=\vert1_20_3\rangle
\end{equation}
This time our light quantum is definitely in the upper arm of the
interferometer, however it is a simple exercise to show that with
this modification no interference can be expected from our MZI.

Delayed choice can be added in the spirit of Wheeler \cite{Whe78},
where the decision to insert or not the second beam splitter is
delayed so that there is a space-like separation between the
passage of our light quantum through $\text{BS}_1$ and the
insertion/removal of $\text{BS}_2$. Experimental results are
unambiguous \cite{Jac07,Hel87,Bal89}: there is no difference
between ``normal'' and ``delayed choice'' modes, therefore there
can be no decision taken beforehand by out light quantum.
One can even continuously morph between ``wave'' and ``particle''
behavior \cite{Ion11}, results \cite{Kai12,Per12} impossible to
explain with classical or hidden variable theories.

One could still speculate if there is a way of spying on the
photons without disturbing them. The answer is positive and in the
following we modify the experiment from
Fig.~\ref{fig:Simple_MZIs_delayed_choice} and show that this
action -- spying on the photons -- is realistic and realizable.

\section{Spying on photons with photons}
\label{sec:spying_on_photons} The experimental setup from
Fig.~\ref{fig:Simple_MZIs_delayed_choice} is modified in the
following way: the mirrors $M_1$ and $M_2$ are replaced by the
beam splitters denoted $\text{BS}_3$ and $\text{BS}_4$. Therefore,
photons may ``leak'' through these beam splitters (by
transmission), bringing therefore with them (under certain
circumstances) the which-path information. To make the experiment
more versatile, these ``leaked'' photons can be brought together
into a new beam splitter, denoted $\text{BS}_5$. Instead of
inserting or not with a delayed choice this beam splitter, we
assume that its (transmission and reflection) parameters can be
modified at will.

\begin{figure}
\centering
\includegraphics[width=3in]{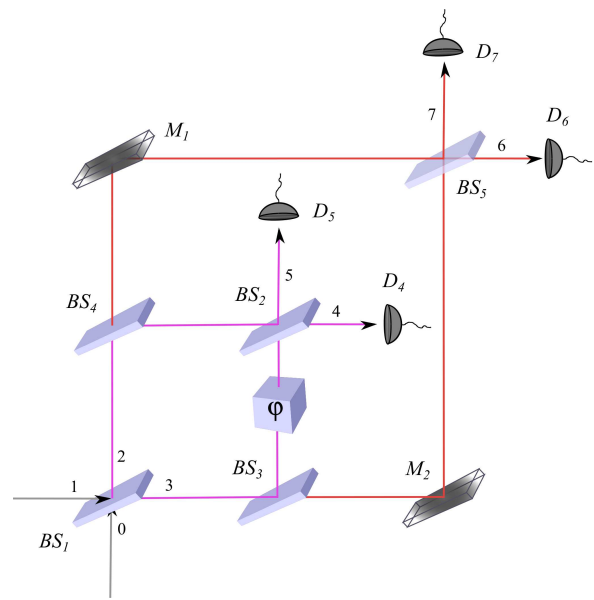}
\caption{Replacing the two mirrors from
Fig.~\ref{fig:Simple_MZIs_delayed_choice} with beam splitters so
that photons can ``leak'' outside, brings us to the new
experimental setup. The outer MZI has the beam splitter
$\text{BS}_5$ characterized by the transmission (reflection)
coefficient $T_w$ ($R_w$). These parameters can be chosen in a
delayed-choice manner.}
\label{fig:braced_MZIs_wave_nature_experiment}
\end{figure}

We end up with the experimental setup depicted in
Fig.~\ref{fig:braced_MZIs_wave_nature_experiment}. This time, at
the two inputs of the beam splitter $\text{BS}_1$ we apply the
state
\begin{eqnarray}
\label{eq:psi_input_state_11}
\vert\psi_{in}\rangle=\vert1_01_1\rangle=\hat{a}_0^\dagger\hat{a}_1^\dagger\vert0\rangle
\end{eqnarray}
i.e. two simultaneously impinging single-photon Fock states.
%
%
If the beam splitter $\text{BS}_1$ is balanced, the wave vector at
its output is
\begin{eqnarray}
\label{eq:psi_after_BS1_state_20_02}
\vert\psi_{23}\rangle=\frac{i}{\sqrt{2}}\vert2_20_3\rangle+\frac{i}{\sqrt{2}}\vert0_22_3\rangle
\end{eqnarray}
This fact is central to the whole experiment. Equation
\eqref{eq:psi_after_BS1_state_20_02} implies that if, for example,
one photon is detected in the lower arm, then \emph{with
certainty} the other one is in the same arm, too. Therefore, one
photon can be forced to interfere with itself while the other one
can be used as which-path marker.

Beam splitters $\text{BS}_1$ through $\text{BS}_4$ are assumed to
be balanced. The beam splitter $\text{BS}_5$, is characterized by
the transmission (reflection) coefficients $T_w$ ($R_w$).
Moreover, we assume that these coefficients can be chosen at any
moment (delayed choice). For example, if $T_w=1$ ($T_w=0$) we have a
configuration where the path from the mirror $M_1$ leads directly
to the detector $D_{6}$ ($D_{7}$). We have therefore maximum which-path knowledge. If $T_w=1/\sqrt{2}$, a photon detected at $D_{6}$ (or $D_7$) could have come with equal likelihood from any path, therefore we have zero which-path information.

Starting from the input state \eqref{eq:psi_input_state_11} and applying the field operator transformations (see details in Appendix
\ref{sec:app:field_oper_transf}) takes us to the output state vector
\begin{eqnarray}
\label{psi_out_is_psi_inner_cross_outer}
\vert\psi_{out}\rangle=\vert\psi_{inner}\rangle+\vert\psi_{cross}\rangle+\vert\psi_{outer}\rangle
\end{eqnarray}
where $\vert\psi_{inner}\rangle$ is the part of the wavevector
corresponding to both photons being inside the inner MZI,
$\vert\psi_{cross}\rangle$ corresponds to the part of the
wavevector where one photon is inside the inner MZI and the
other one inside the outer one etc. The interesting part is found
in $\vert\psi_{cross}\rangle$ and we have
\begin{eqnarray}
\label{eq:psi_out_cross_11_input} \vert\psi_{cross}\rangle=
-\frac{T_w+iR_w\text{e}^{i\varphi}}{2\sqrt{2}}\vert1_41_{6}\rangle
-\frac{iT_w\text{e}^{i\varphi}+R_w}{2\sqrt{2}}\vert1_41_{7}\rangle
\nonumber\\
-\frac{iT_w+R_w\text{e}^{i\varphi}}{2\sqrt{2}}\vert1_51_{6}\rangle
-\frac{T_w\text{e}^{i\varphi}+iR_w}{2\sqrt{2}}\vert1_51_{7}\rangle\quad
\end{eqnarray}
One could focus on one of the coincidence detections \emph{e.g.} at
detectors $D_4$ and $D_6$,
\begin{eqnarray}
\label{eq:P_46_without_epsilon}
P_{4-6}=\vert\langle1_41_6\vert\psi_{out}\rangle\vert^2=\vert\langle1_41_6\vert\psi_{cross}\rangle\vert^2.
\end{eqnarray}
We shall introduce a parameter $\varepsilon\in[0,1]$
characterizing the beam splitter $\text{BS}_5$ and we make the choice
$T_w=\varepsilon$ and $R_w=i\sqrt{1-\varepsilon^2}$. Therefore,
after a short computation equation \eqref{eq:P_46_without_epsilon} becomes
\begin{eqnarray}
\label{eq:P_46_epsilon}
P_{4-6}=\frac{1-2\varepsilon\sqrt{1-\varepsilon^2}\cos\left(\varphi\right)}{8}
\end{eqnarray}
For $T_w=0$ (or $T_w=1$) the which-path information of the inner
photon is known \emph{with certainty} through the outer one
detected at $D_6$. The coincidence probability now gives
\begin{equation}
\label{eq:P4_6_particle-like_1_over_8}
P_{4-6}=\frac{1}{8}
\end{equation}
As expected, all interference is gone. On the
other hand maximum uncertainty on the path taken by the photon in the inner MZI is
obtained when $\text{BS}_5$ is balanced ($\varepsilon=1/\sqrt{2}$)
yielding in this case 
\begin{equation}
P_{4-6}=\frac{1-\cos\left(\varphi\right)}{8}
\end{equation}
i.e. interference fringes with 100\% visibility.
For other values of $\varepsilon$ one has partial information
about the path quantified by
\begin{eqnarray}
K=\big\vert\vert{T_w}\vert^2-\vert{R_w}\vert^2\big\vert=\vert2\varepsilon^2-1\vert
\end{eqnarray}
and interference fringes with a visibility given by
\begin{eqnarray}
V=2\vert{T_w}\vert\vert{R_w}\vert=2\varepsilon\sqrt{1-\varepsilon^2}.
\end{eqnarray}
It is easy to show that we have
\begin{eqnarray}
\label{eq:C_sq_plus_V_sq_is_1} K^2+V^2=1
\end{eqnarray}
for all values of $\varepsilon$ and we obtained again the extreme case of the well-known inequality
quantifying the duality between which-path information and
interference \cite{Woo79,Dur00,Gre88,Eng95,Eng96}.

\begin{figure}
\centering
\includegraphics[width=3in]{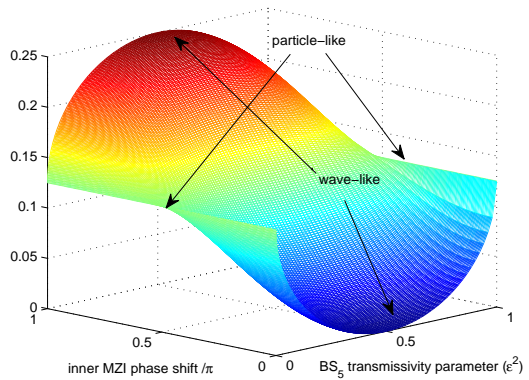}
\caption{The coincidence probability at $D_4$ and $D_{6}$ as a
function of the path length difference $\varphi$ and the ``delayed
choice'' beam splitter $\text{BS}_5$ parameter $\varepsilon^2$. As
$\varepsilon$ goes from $0$ through $1/\sqrt{2}$ to $1$ we go from
maximum particle-like through maximum wave-like and back again to
maximum particle-like behavior.} \label{fig:Pc_4_6}
\end{figure}

The coincidence counts probability $P_{4-6}\left(\varphi,\varepsilon\right)$ at detectors $D_4$ and $D_6$
is plotted in Fig.~\ref{fig:Pc_4_6}. As it can be seen, we can continuously
morph from ``particle-like'' to ``wave-like'' behavior by
modifying the transmission parameter $\varepsilon$ of the beam
splitter $\text{BS}_5$.

The interesting point about this experiment is that the detection
event at $D_4$ can take place much earlier than the detection
event at $D_{6}$, one can even make these event space-like
separated, therefore any interpretation of pilot waves or other
signals travelling both routes in the inner interferometers are
simply not tenable. The decision of what aspect (wave, particle or
a little of both) to watch is taken much later, long after the
photon in the inner path has been detected and destroyed. The
decision what to measure is in the hands of the experimenter.

It is noteworthy that ignoring the
inner detectors and focusing on singles detections at, say, $D_6$
or $D_7$ will show no interference (see computational details in
Appendix \ref{sec:app:matrix_density_approach}).

\section{Varying the which-path information with a modified input state}
\label{sec:spying_modif_input_state}
The previous section ended with the conclusion that spying on the
photon from the inner MZI with $100\%$ certainty prevented us to observe
interference effects. As stated earlier, the crucial point in our
experiment was given by equation
\eqref{eq:psi_after_BS1_state_20_02}. Namely, we had absolute
certainty that if one photon is in one arm of the interferometer,
the other one will be there, too.

There is another state that is interesting in this respect, namely
the input state
\begin{eqnarray}
\label{eq:psi_in_prime}
\vert\psi'_{in}\rangle=\frac{1}{\sqrt{2}}\left(\vert2_00_1\rangle+\vert0_02_1\rangle\right)
\end{eqnarray}
yielding after the beam splitter $\text{BS}_1$ the state vector
\begin{eqnarray}
\label{eq:psi_after_BS1_state_11_prime}
\vert\psi'_{23}\rangle=i\vert1_21_3\rangle
\end{eqnarray}
This time we also have absolute certainty that if one photon is
detected in one arm, the other one will be in the \emph{opposite} arm. Using
the same principle as before, we find
\begin{eqnarray}
\vert\psi'_{out}\rangle=\vert\psi'_{inner}\rangle+\vert\psi'_{cross}\rangle+\vert\psi'_{outer}\rangle
\end{eqnarray}
where $\vert\psi'_{cross}\rangle$ is given by
\begin{eqnarray}
\label{eq:psi_out_cross_2002_input} \vert\psi'_{cross}\rangle=
-\frac{iT_w\text{e}^{i\varphi}+R_w}{2\sqrt{2}}\vert1_41_{6}\rangle
-\frac{T_w+iR_w\text{e}^{i\varphi}}{2\sqrt{2}}\vert1_41_{7}\rangle
\nonumber\\
-\frac{T_w\text{e}^{i\varphi}+iR_w}{2\sqrt{2}}\vert1_51_{6}\rangle
-\frac{iT_w+R_w\text{e}^{i\varphi}}{2\sqrt{2}}\vert1_51_{7}\rangle\quad
\end{eqnarray}
Computing again the coincidence probability
\begin{eqnarray}
P_{4-6}=\vert\langle1_41_6\vert\psi'_{out}\rangle\vert^2=\vert\langle1_41_6\vert\psi'_{cross}\rangle\vert^2
\end{eqnarray}
takes us to
\begin{eqnarray}
\label{eq:P_46_epsilon_prime}
P_{4-6}=\frac{1+2\varepsilon\sqrt{1-\varepsilon^2}\cos\left(\varphi\right)}{8}
\end{eqnarray}
The same conclusions from the previous section apply. One could therefore
conclude that we have a ``particle-like'' setting, namely for
$T_w=0$ or $T_w=1$ yielding again the result from equation \eqref{eq:P4_6_particle-like_1_over_8} and a ``wave-like'' setting for
$T_w=1/\sqrt{2}$ yielding
\begin{eqnarray}
\label{eq:P_46_psi_prime}
P_{4-6}=\frac{1+\cos\left(\varphi\right)}{8}
\end{eqnarray}

In the following, we will show that this is
actually an \emph{incomplete picture}. By simply changing the input
state, we shall render useless this classification.

The state vector $\vert\psi_{23}\rangle$ from equation
\eqref{eq:psi_after_BS1_state_20_02} guaranteed us that the two
photons are \emph{always} in the same arm of the interferometer
right after $\text{BS}_1$ while $\vert\psi'_{23}\rangle$ from
equation \eqref{eq:psi_after_BS1_state_11_prime} guarantees us
that the two photons are \emph{never} in the same arm. We could
erase (partially or totally) this information by preparing an
input state that is a coherent superposition of being both in
$\vert\psi_{in}\rangle$ and in $\vert\psi'_{in}\rangle$.
Therefore, we define
\begin{eqnarray}\label{eq:psi_input_state_sec}
\vert\psi''_{in}\rangle=\alpha\vert\psi_{in}\rangle+\sqrt{1-\alpha^2}\vert\psi'_{in}\rangle
\end{eqnarray}
and after the beam splitter $\text{BS}_1$ we have the state vector
\begin{eqnarray}
\label{eq:psi_after_BS1_state_alpha}
\vert\psi''_{23}\rangle=\alpha\vert\psi_{23}\rangle+\sqrt{1-\alpha^2}\vert\psi'_{23}\rangle
\end{eqnarray}
with $\alpha\in\lbrack0,1\rbrack$.
Using the same technique as before we can compute the output state
vector $\vert\psi''_{out}\rangle$. One can obtain now the
coincidence probability at the detectors $D_4$ and $D_6$ as
\begin{eqnarray}
\label{eq:P_46_epsilon_sec_general} P_{4-6}=
\vert\langle1_41_6\vert\psi''_{out}\rangle\vert^2=\vert\langle1_41_6\vert\psi''_{cross}\rangle\vert^2
\nonumber\\
=\frac{1-2\alpha\sqrt{1-\alpha^2}\sin\left(\varphi\right)+2\left(1-2\alpha^2\right)\varepsilon\sqrt{1-\varepsilon^2}\cos\left(\varphi\right)}{8}
\quad
\end{eqnarray}
where $\vert\psi''_{cross}\rangle=\alpha\vert\psi_{cross}\rangle+\sqrt{1-\alpha^2}\vert\psi'_{cross}\rangle$. We have now the path-related
information
\begin{equation}
K=\vert2\alpha^2-1\vert\big\vert\vert{T_w}\vert^2-\vert{R_w}\vert^2\big\vert
=\vert2\alpha^2-1\vert\vert2\varepsilon^2-1\vert
\end{equation}
and the visibility of the interference pattern
\begin{eqnarray}
V=2\sqrt{\alpha^2\left(1-\alpha^2\right)+\left(2\alpha^2-1\right)^2\vert{T_w}\vert^2\vert{R_w}\vert^2}
\nonumber\\
=2\sqrt{\alpha^2\left(1-\alpha^2\right)+\left(2\alpha^2-1\right)^2\varepsilon^2\left(1-\varepsilon^2\right)}
\end{eqnarray}
One can easily check that equation \eqref{eq:C_sq_plus_V_sq_is_1}
is again satisfied for any $\alpha$ and $\varepsilon$. The least which-path information about the inner MZI photon that
can be harvested by a detection with a ``leaked'' photon at $D_6$
(or $D_7$) from the state $\vert\psi''_{in}\rangle$ is obtained
for $\alpha=1/\sqrt{2}$. In this case the coincidence count
probability given by equation \eqref{eq:P_46_epsilon_sec_general} becomes
\begin{eqnarray}
\label{eq:P_46_epsilon_sec}
P_{4-6}=\frac{1-\sin\left(\varphi\right)}{8}
\end{eqnarray}
It is worthwhile to note that this time the coincidence
probability is wave-like and does not depend on $T_w$ ($R_w$).
This can be coupled by the fact that no which-path information can
be extracted from equation \eqref{eq:psi_input_state_sec} when
$\alpha=1/\sqrt{2}$ (as proven by the fact that $K=0$ in this
case). Therefore, is it quite remarkable that the photon from the
inner interferometer is willing to show us an interference pattern
with $100\%$ visibility, no matter if we insert or not the beam
splitter $\text{BS}_5$. It is equally remarkable that in this case the single detection probability $P_4$ at the detector $D_4$ shows the same behavior.

\begin{figure}
\centering
\includegraphics[width=3in]{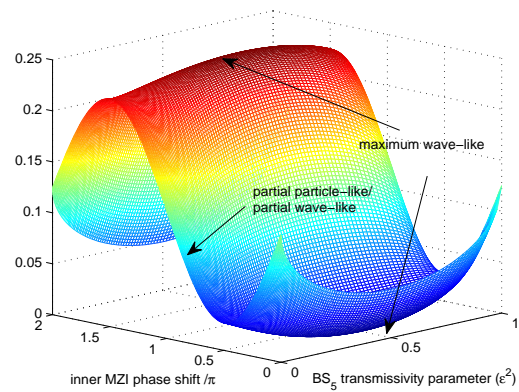}
\caption{The coincidence counts probability $P_{4-6}$ at $D_4$ and $D_{6}$ as a
function of the path length difference $\varphi$ and the beam
splitter $\text{BS}_5$ parameter $\varepsilon^2$ for $\alpha=0.9$.
There is no more a clear ``particle-like'' behavior as in
Fig.~\ref{fig:Pc_4_6}. } \label{fig:Pc_4_6_alpha_0_2}
\end{figure}

In Fig.~\ref{fig:Pc_4_6_alpha_0_2} we plot the coincidence counts
probability $P_{4-6}$ from equation
\eqref{eq:P_46_epsilon_sec_general} for $\alpha=0.9$. Since less
which-path information can be now extracted from the detection at
$D_6$ (we have $K=0.62\vert2\varepsilon^2-1\vert$) the visibility
of the interference fringes is quite high. We can have a perfect
``wave-like'' behavior for $\varepsilon=1/\sqrt{2}$ (when $K=0$
and $V=1$), the worst case being for $\varepsilon=0$ (or
$\varepsilon=1$) when $V=0.78$.

\section{Conclusions}
\label{conclusions} In this paper reconsidered the so-called
wave-particle duality from a different point of view. We extended
the much-discussed simple Mach-Zehnder interferometer into two
braced interferometers. The ability to convey information about
the path taken by one photon immediately limits its wave-like
behavior, as discussed in Section \ref{sec:spying_on_photons}. Contrary to all proposed and performed experiments up-to-date, we show that it is possible to have a constant
``wave-like'' behavior of photon from the inner MZI no
matter if the beam splitter $\text{BS}_5$ is inserted or not by simply engineering
our input state vector. 
We conclude that there are no pre-determined ``wave-like'' or ``particle-like'' experimental setups, it all boils down to how much information we can
extract from our system.


\appendix



\section{Computation of the field operator transformations}
\label{sec:app:field_oper_transf} In the general case, when the
beam splitters are not balanced, we have the field operator
transformations (see, e. g. \cite{Ata14b} for a similar
computation):
\begin{eqnarray}
\label{eq:a0_dagger_transformation}
\hat{a}_0^\dagger=R\left(T^2+R^2\text{e}^{i\varphi}\right)\hat{a}_4^\dagger+TR^2\left(1+\text{e}^{i\varphi}\right)\hat{a}_5^\dagger
\nonumber\\
+T\left(TT_w+RR_w\right)\hat{a}_{6}^\dagger+T\left(TR_w+RT_w\right)\hat{a}_{7}^\dagger
\end{eqnarray}
and
\begin{eqnarray}
\label{eq:a1_dagger_transformation}
\hat{a}_1^\dagger=TR^2\left(1+\text{e}^{i\varphi}\right)\hat{a}_4^\dagger+R\left(T^2\text{e}^{i\varphi}+R^2\right)\hat{a}_5^\dagger
\nonumber\\
+T\left(TR_w+RT_w\right)\hat{a}_{6}^\dagger+T\left(TT_w+RR_w\right)\hat{a}_{7}^\dagger
\end{eqnarray}
For the balanced case, the state operator transformations can be
written as
\begin{eqnarray}
\label{eq:a0_dagger_transformation_balanced}
\hat{a}_0^\dagger=\frac{\text{e}^{i\frac{\varphi}{2}}\sin\left(\frac{\varphi}{2}\right)}{\sqrt{2}}\hat{a}_4^\dagger
-\frac{\text{e}^{i\frac{\varphi}{2}}\cos\left(\frac{\varphi}{2}\right)}{\sqrt{2}}\hat{a}_5^\dagger
\nonumber\\
+\frac{T_w+iR_w}{2}\hat{a}_{6}^\dagger+\frac{iT_w+R_w}{2}\hat{a}_{7}^\dagger
\end{eqnarray}
and
\begin{eqnarray}
\label{eq:a1_dagger_transformation_balanced}
\hat{a}_1^\dagger=-\frac{\text{e}^{i\frac{\varphi}{2}}\cos\left(\frac{\varphi}{2}\right)}{\sqrt{2}}\hat{a}_4^\dagger
-\frac{\text{e}^{i\frac{\varphi}{2}}\sin\left(\frac{\varphi}{2}\right)}{\sqrt{2}}\hat{a}_5^\dagger
\nonumber\\
+\frac{iT_w+R_w}{2}\hat{a}_{6}^\dagger+\frac{T_w+iR_w}{2}\hat{a}_{7}^\dagger
\end{eqnarray}
Taking now the input state \eqref{eq:psi_input_state_11} and using
the field operator transformations
\eqref{eq:a0_dagger_transformation_balanced} and
\eqref{eq:a1_dagger_transformation_balanced} we end up with the
state vector given by equation
\eqref{psi_out_is_psi_inner_cross_outer} where
$\vert\psi_{cross}\rangle$ is given by equation
\eqref{eq:psi_out_cross_11_input} and we have
\begin{eqnarray}
\label{eq:psi_out_inner_11_state} \vert\psi_{inner}\rangle
=\frac{\text{e}^{i\varphi}\sin\left(\varphi\right)}{2\sqrt{2}}
\left(\vert0_42_5\rangle-\vert2_40_5\rangle\right)
\nonumber\\
-\frac{\text{e}^{i\varphi}\cos\left(\varphi\right)}{2}\vert1_41_5\rangle
\end{eqnarray}
and
\begin{eqnarray}
\label{eq:psi_out_outer_11_state}
\vert\psi_{outer}\rangle=i\frac{T_w^2+R_w^2}{2\sqrt{2}}\left(\vert2_{6}0_{7}\rangle
+\vert0_{6}2_{7}\rangle\right)
+iT_wR_w\vert1_{6}1_{7}\rangle\quad
\end{eqnarray}

\section{Ignoring some of the detectors -- the density matrix approach}
\label{sec:app:matrix_density_approach} If we chose to ignore
everything that happens outside the inner interferometer, then we
have to work in the density matrix approach. Therefore, we first
construct the global density matrix
$\hat{\rho}_{out}=\vert\psi_{out}\rangle\langle\psi_{out}\vert$
and trace over the unused outputs (in this case $6$ and $7$),
yielding the reduced density matrix
\begin{eqnarray}
\label{eq:rho_out_trace_10_11}
\hat{\rho}_{4-5}=\text{Tr}_{6,7}\left\{\hat{\rho}_{out}\right\}=\sum_{m,n\in\mathbb{N}}{\langle
m_{6}n_{7}\vert\hat{\rho}_{out}\vert m_{6}n_{7}\rangle}
\end{eqnarray}
The central question is now if one could see an interference
pattern while selecting only single counts (i.e. one and only one
detection event at either $D_4$ or $D_5$) in the inner
interferometer while completely ignoring what happens outside it.
It is straightforward to check that this is not the case. For
example, the single count rate at detector $D_4$ is
\begin{eqnarray}
\label{eq:Psingle_D6}
P_{4}=\text{Tr}\left\{\hat{a}_4^\dagger\hat{a}_4\hat{\rho}_{4-5}\right\}
=\frac{1}{2}
\end{eqnarray}
The reader can easily check that we get an identical answer by
imposing one photo-count at $D_5$ and none at $D_4$ (and still
ignoring the outer detectors).



\begin{thebibliography}{6}


\bibitem{Boh49}
N. Bohr, \emph{Discussions with Einstein on Epistemological
Problems in Atomic Physics}, Cambridge University Press (1949)

\bibitem{Fey65}
R. P. Feynman, R. B. Leighton, and M. Sands, \emph{The Feynman
Lectures on Physics} (Addison-Wesley, Reading, MA, 1965), Vol.
III, Chap. 1.

\bibitem{Boh28}
N. Bohr, 
Naturwissenschaften \textbf{16}, 245 (1928)






\bibitem{Woo79}
W. Wootters, W. H. Zurek, 
Phys. Rev. D \textbf{19}, 473 (1979)

\bibitem{Dur00}
S. D\"urr, G. Rempe, 
Am. J. Phys. \textbf{68} , 1021 (2000)


\bibitem{Gre88}
D.M. Greenberger, A. Yasin, 
Phys. Lett. A \textbf{128}, 391 (1988)


\bibitem{Eng95}
B.-G. Englert, M. O. Scully, H. Walther, 
Nature \textbf{375}, 367 (1995)

\bibitem{Eng96}
B.-G. Englert, 
Phys. Rev. Lett. \textbf{77}, 2154 (1996)


\bibitem{Whe78}
J. Wheeler, \emph{Problems in Formulation of Physics}, ed. G. t.
di Francia, (North-Holland, Amsterdam, 1978)

\bibitem{Whe83}
J. Wheeler's ``Law without law'' in \emph{Quantum Theory and
Measurement}, edited by J. A. Wheeler and W. H. Zurek (Princeton
University Press, Princeton, NJ, 1983)


\bibitem{Jac07}
V. Jacques \emph{et al.} 
Science \textbf{315}, 966 (2007)

\bibitem{Jac08}
V. Jacques \emph{et. al.}, 
New J. Phys. \textbf{10}, 123009 (2008)


\bibitem{Hel87}
T. Hellmuth, H. Walther, A. Zajonc, W. Schleich, 
Phys. Rev. A \textbf{35}, 2532 (1987);

\bibitem{Bal89}
J. Baldzuhn, E. Mohler, W. Martienssen, 
Z. Phys. B \textbf{77}, 347 (1989).




\bibitem{Cha95}
M. S. Chapman  \emph{et al.} 
Phys. Rev. Lett. \textbf{75}, 3783 (1995).

\bibitem{Eic93}
U. Eichman \emph{et al.} 
Phys. Rev. Lett. \textbf{70}, 2359 (1993).

\bibitem{Dur98}
S. D\"urr, T. Nonn, G. Rempe, 
Nature \textbf{395}, 33 (1998).

\bibitem{Ber01}
P. Bertet \emph{et al.} 
Nature \textbf{411}, 10 (2001)





\bibitem{Ion11}
R. Ionicioiu, D. R. Terno, 
Phys. Rev. Lett. \textbf{107}, 230406 (2011)


\bibitem{Kai12}
F. Kaiser et al., 
Science \textbf{338}, 637 (2012)


\bibitem{Per12}
A. Peruzzo et al., 
Science \textbf{338}, 634 (2012)



\bibitem{Scu82}
M. O. Scully, K. Dr\"uhl, 
Phys. Rev. A \textbf{25}, 2208 (1982)

\bibitem{Scu91}
M. O. Scully, B.-G. Englert, H. Walther, 
Nature \textbf{351}, 111 (1991)

\bibitem{Eng99}
B.-G. Englert, M. O. Scully, H. Walther, 
Am. J. Phys. \textbf{67}, 4, (1999) 

\bibitem{Scu98}
M. O. Scully, H. Walther, 
Found. Phys. \textbf{28}, 399 (1998)

\bibitem{Wal02}
S. P. Walborn, M. O. Terra Cunha, S. Padua, C. H. Monken, 
Phys. Rev. A \textbf{65}, 033818 (2002)

\bibitem{Hil98}
R. Hilmer, P. G. Kwiat, 
Sci. Am. \textbf{296}, 90 (1998)

\bibitem{Ma16}
X. Ma, J. Kofler, A. Zeilinger, 
Rev. Mod. Phys. \textbf{88}, 015005 (2016)

\bibitem{Ou97}
Z. Y. Ou, 
Phys. Lett. A \textbf{226}, 323 (1997)

\bibitem{Ou90}
Z. Y. Ou et al., 
Phys. Rev. A \textbf{41}, 566 (1990)



\bibitem{Heu15}
A. Heuer, R. Menzel, P. W. Milonni, 
Phys. Rev. Lett. \textbf{114}, 053601 (2015)

\bibitem{Ata14b}
S. Ataman, 
Eur. Phys. J. D \textbf{68}, 317 (2014)




\bibitem{Lou03}
R. Loudon, \emph{The Quantum Theory of Light}, (Oxford University
Press, Third Edition, 2003)

\bibitem{Enk05}
S. J. van Enk, 
Phys. Rev. A \textbf{72} , 064306 (2005)

\bibitem{Ata14a}
S. Ataman, 
Eur. Phys. J. D \textbf{68}, 288 (2014)




\end{thebibliography}
\end{document}